\documentclass{article}
\usepackage{frascatiphys}
\usepackage{graphicx}
\begin{document}
\title{ 
THE KOTO II EXPERIMENT
}
\author{
Silvia Martellotti        \\
{\em INFN Laboratori Nazionali di Frascati} \\
}
\maketitle
\baselineskip=11.6pt
\begin{abstract}
$K_L \rightarrow \pi^0\nu\bar\nu$ is a golden mode for flavour physics: it is highly suppressed in the Standard Model, with a predicted branching ratio of $3 \times 10^{-11}$ and a theoretical uncertainty of just 2\%, and is therefore extremely sensitive to new physics. The KOTO experiment at J-PARC is searching for this decay. Using the data taken in 2021, it has set the most stringent upper limit to date of $2.2 \times 10^{-9}$ on the branching ratio at 90\% C.L. We present KOTO II, a next-generation experiment at J-PARC proposed in 2024, which aims to measure the $K_L \rightarrow \pi^0\nu\bar\nu$ branching ratio with a sensitivity below $10^{-12}$ in the 2030s.
\end{abstract}
\baselineskip=14pt

\section{Introduction}
The rare kaon decay $K_L \rightarrow \pi^0\nu\bar\nu$ is extremely sensitive to new physics, because it is highly suppressed in the Standard Model (SM) and very precisely predicted: the SM predicts the branching fraction of this decay to be $(2.94 \pm 0.15) \times 10^{-11}$\cite{buras}, with a theoretical uncertainty of just 2\%. Measuring this branching ratio could provide essential new information about the flavour structure of the quark sector through the $s \rightarrow d$ transition.
The decay is being searched for with the KOTO experiment at J-PARC, Japan. The current best upper limit on its branching ratio is $2.2 \times 10^{-9}$ at 90\% C.L., obtained with KOTO data taken in 2021. KOTO is expected to reach a sensitivity below $10^{-10}$ by the end of the decade, while the SM prediction is of the orther of $O(10^{-11})$. A next-generation experiment, KOTO II, was proposed in 2024 with the goal of observing the decay with more than $5\sigma$ significance in the 2030s at the SM value of the branching ratio. The proposal was prepared by 82 members from 11 countries. The observation of the decay $K_L\rightarrow \pi^0 \ell^+\ell^-$ is another important aim of KOTO II, together with searches for other rare decays and hidden-sector particles. After 2026, KOTO will be the only running experiment dedicated to rare kaon decays, and KOTO II is currently the only planned future kaon experiment in the world. The KOTO II Collaboration is intended as a global effort in kaon physics, with significant contributions and support from the European community.

\section{KOTO experiment}

\subsection{Experimental apparatus}
\begin{figure}[htb]
    \begin{center}
        {\includegraphics[width=1\textwidth]{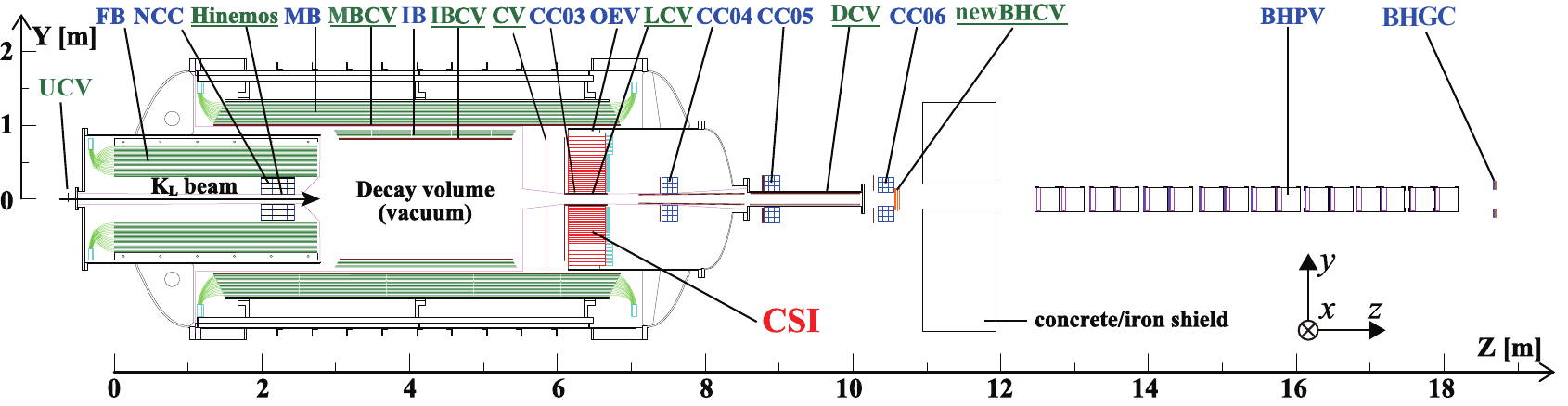}}
        \caption{\it Cross-sectional view of the KOTO detector. The beam enters from the left. Detector components with their abbreviated names written in blue regular (in green regular and underlined) are photon (charged-particle) veto counters.}
\label{koto_schematic}
    \end{center}
\end{figure}

A 30 GeV proton beam from the J-PARC main ring is extracted for 2 s every 5.2 s towards the gold target \cite{koto_beam} located in the Hadron Experimental Facility. The particles produced in the target are transported into the KOTO detector at an angle of $16^\circ$ from the proton beam via a 20 m long neutral beam line. The beam line consists of two collimators and a sweeping magnet between them \cite{koto_mag}. The collimators were aligned with a beam profile monitor \cite{monitor} to minimize beam particles outside the nominal beam size due to interactions with beam line components, called ``beam-halo'' particles. The solid angle of the beam is $7.8~\mu$sr, and its size is $8\times8$ cm$^2$ at the end of the downstream collimator. The neutral beam is composed of $K_L$'s, neutrons, and photons. The $K_L$ flux, with a momentum distribution peaking at $1.4$ GeV$/c$, was measured to be $2.1 \times 10^{-7}$ $K_L$'s per proton on target (POT). The cross-sectional view of the KOTO detector is shown in Fig.~\ref{koto_schematic}. The $z$ axis is along the beam direction, and the origin is set at the upstream edge of FB, 21.5 m from the target.

The $K_L \rightarrow \pi^0\nu\bar\nu$ decay is identified by detecting the two photons from the $\pi^0$ decay while observing nothing else in hermetic veto counters around the decay region in vacuum and downstream. A 2 m diameter cylindrical electromagnetic calorimeter (CSI) \cite{calo} was centered along the beam axis with a $15 \times 15$ cm$^2$ beam hole to measure the energies and positions of the two photons. CSI consisted of 2716 undoped CsI crystals, 50 cm long, with a cross section of $2.5 \times 2.5$ cm$^2$ ($5 \times 5$ cm$^2$) inside (outside) the central $1.2 \times 1.2$ m$^2$ region. In front of CSI, two layers of plastic scintillating counters (CV) \cite{CV} were placed to identify charged particles entering CSI. The decay region was surrounded by lead/scintillator sampling calorimeters, IB, MB, FB \cite{IB,FB}, and OEV \cite{OEV}. NCC, CC03, CC04, CC05, and CC06, composed of undoped CsI crystals, encircled the beam line to detect particles escaping along the beam direction. Behind CC06, three layers of wire chambers (newBHCV) and sixteen aerogel Cherenkov counters with lead converters (BHPV) \cite{BHPV} were placed in the beam core to detect charged particles and photons escaping through the beam hole. The following counters were added since 2018 to further suppress backgrounds. In 2018, Multi-Pixel Photon Counters (MPPCs) were attached at the upstream surface of the CSI crystals to improve neutron identification \cite{MPPC}. A new scintillating counter (DCV) \cite{DCV} was installed on the inner surface of the downstream vacuum pipe behind CSI to detect charged particles passing through the CSI beam hole. In 2021, a charged veto counter consisting of 0.5 mm square scintillating fibers (UCV) \cite{UCV} was installed in the beam core at $z = -920$ mm to detect charged particles before the entrance of the detector.

\subsection{Event reconstruction and selection}
The analyzed data recorded in 2021 correspond to $3.3 \times 10^{19}$ POT, with a proton-beam intensity ranging from 10 to 64 kW.

The photon energy and position in CSI were reconstructed from a cluster of adjacent crystals with deposited energies larger than 3 MeV. The $\pi^0$ vertex position ($Z_{\mathrm{vtx}}$) was reconstructed by assuming the $\pi^0 \rightarrow \gamma\gamma$ decay on the beam axis. The $\pi^0$ momentum perpendicular to the beam direction ($P_t$) was then calculated. The $\pi^0$ in $K_L \rightarrow \pi^0\nu\bar\nu$ decay was required to have large $P_t$ to account for the missing momentum taken by the two neutrinos and to avoid the $K_L \rightarrow \pi^+\pi^-\pi^0$ background, where the $\pi^0$ $P_t$ is limited to below 133 MeV$/c$ by a kinematic boundary. The signal region was defined in the $P_t$ vs $Z_{\mathrm{vtx}}$ plane as $130 < P_t < 250$ MeV$/c$ and $3200 < Z_{\mathrm{vtx}} < 5000$ mm. The $Z_{\mathrm{vtx}}$ range was chosen to avoid $\pi^0$'s generated by interactions of beam-halo neutrons with NCC and CV. Details of the selection criteria are given in \cite{UL_KOTO}.

\subsection{Background estimation}

\begin{figure}[htb]
    \begin{center}
        {\includegraphics[width=.47\linewidth]{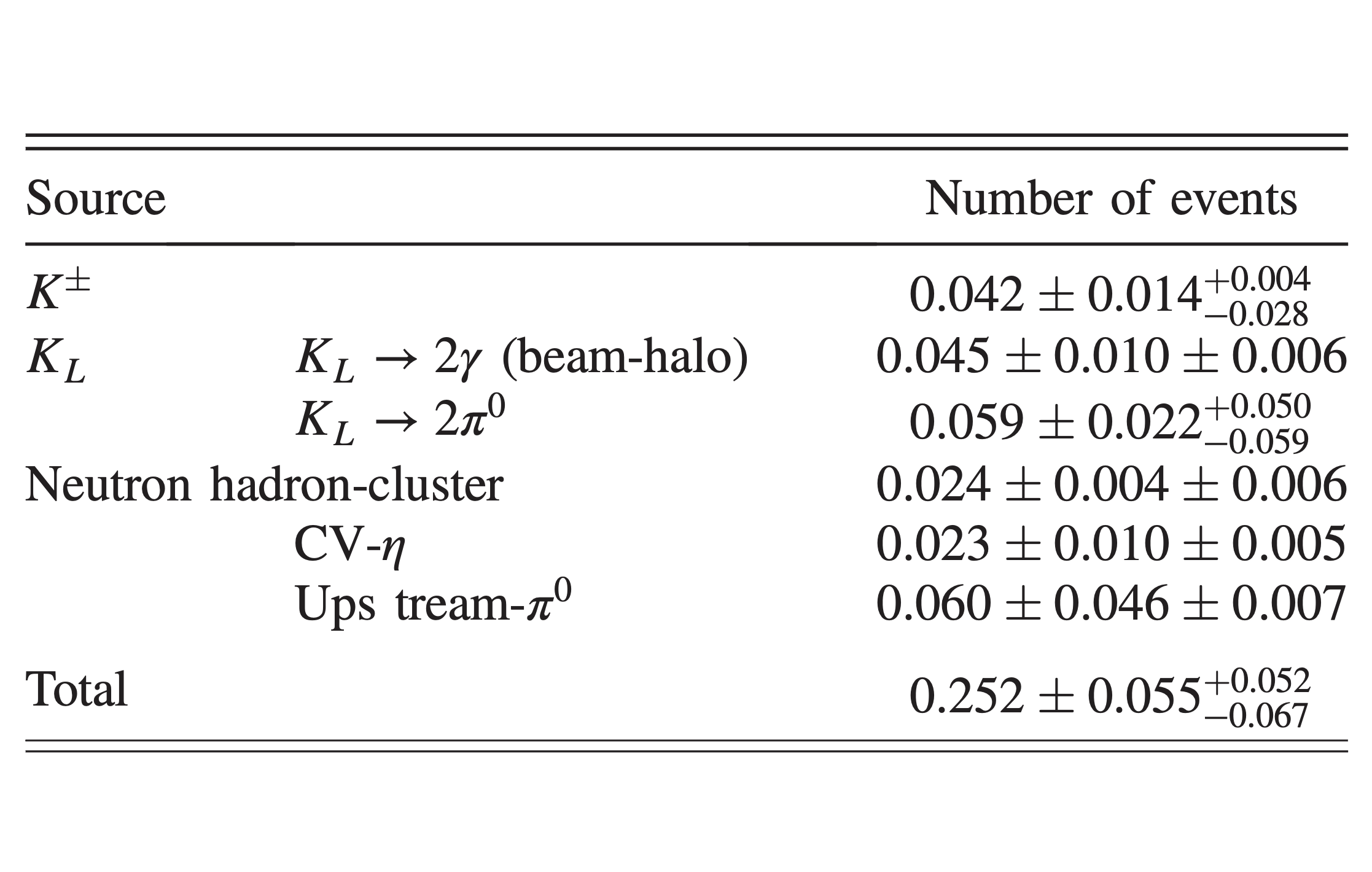}}
\hspace{0.4cm}
        {\includegraphics[width=.47\linewidth]{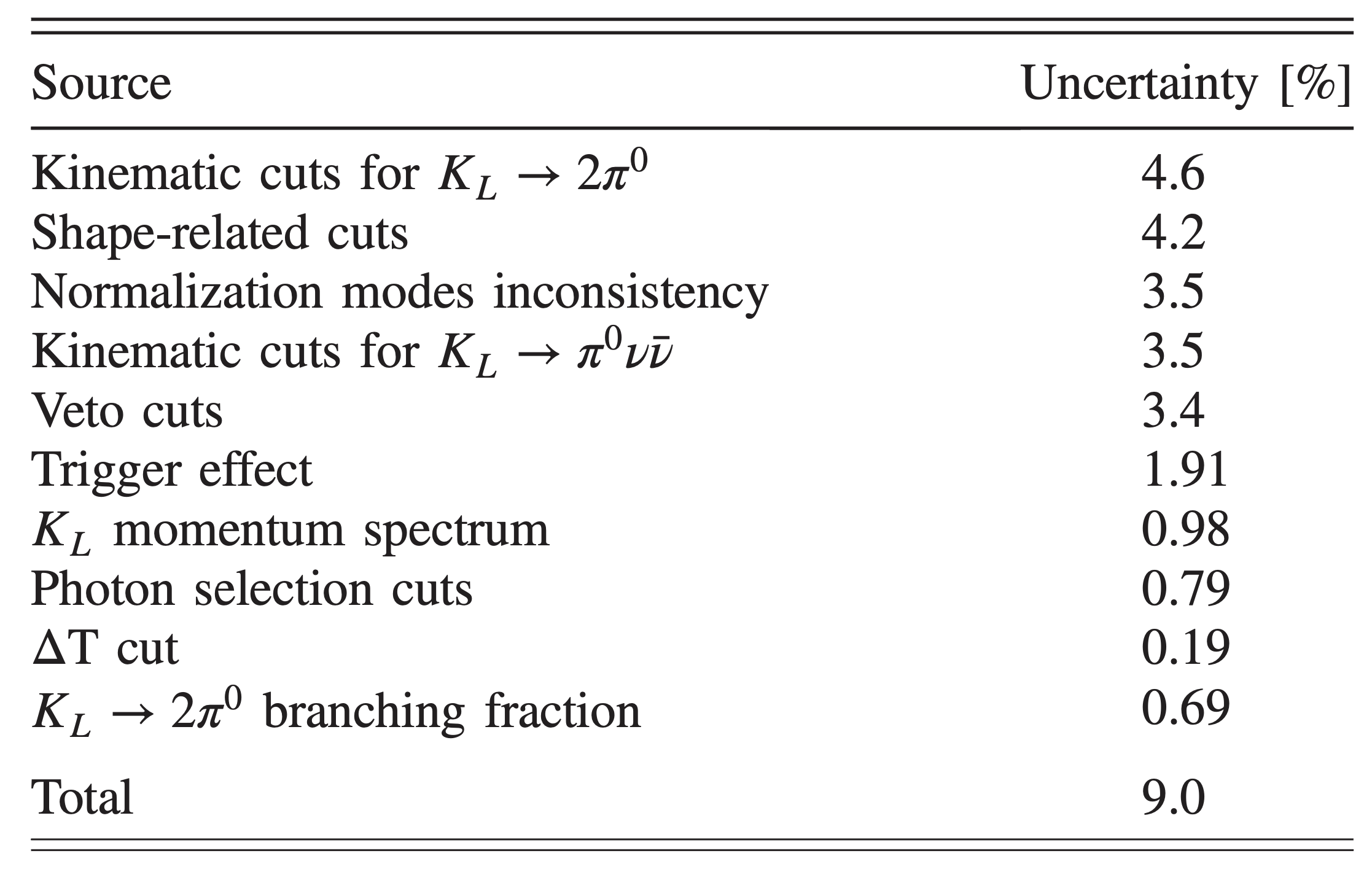}}
        \caption{\it Left: summary of background estimation. The second (third) number represent the statistical uncertainties (systematic uncertainties). Right: relative systematic uncertainties on the single event sensitivity.}
\label{tables}
    \end{center}
\end{figure}

The left table in Fig.~\ref{tables} summarizes the background estimation. The total number of background events in the signal region was estimated to be $0.252\pm 0.055_{\mathrm{stat}}{}^{+0.052}_{-0.067\,\mathrm{syst}}$. Background sources fall into three groups: $K^\pm$ decay backgrounds, $K_L$ decay backgrounds, and neutron-induced backgrounds.
$K^\pm$'s are generated from interactions of $K_L$ particles with the downstream collimator, after the sweeping magnet, and may then enter the KOTO detector. A $\pi^0$ from $K^\pm$ decays may have kinematics similar to that of the $\pi^0$ from $K_L \rightarrow \pi^0\nu\bar\nu$. The $K^\pm$ background was suppressed in the 2021 analysis by detecting $K^\pm$'s with UCV. The $K^\pm$-to-$K_L$ flux ratio at the beam exit was evaluated to be $3.3 \times 10^{-5}$ by identifying $K^\pm \rightarrow \pi^\pm\pi^0$ decays with three-cluster events in CSI. The number of $K^\pm$ background events was estimated to be $0.042\pm 0.014_{\mathrm{stat}}{}^{+0.004}_{-0.028\,\mathrm{syst}}$, based on MC simulation incorporating the UCV inefficiency obtained from data, normalized to data using the measured $K^\pm$-to-$K_L$ ratio.
A beam-halo $K_L$ that decays to $2\gamma$ could have large reconstructed $P_t$ and become a background. The beam-halo $K_L$ flux was estimated by MC simulation normalized to data using $K_L \rightarrow 3\pi^0$ events. The number of beam-halo $K_L \rightarrow 2\gamma$ background events was estimated to be $0.045\pm 0.010_{\mathrm{stat}}\pm 0.006_{\mathrm{syst}}$.
The $K_L \rightarrow 2\pi^0$ background was evaluated by MC simulation, taking into account the difference in photon detection inefficiency of the veto counters between data and MC. The inefficiency of FB, IB, MB, and BHPV was evaluated using $K_L \rightarrow 3\pi^0$ events with five of the six photons hitting CSI. The number of $K_L \rightarrow 2\pi^0$ background events was estimated to be $0.059\pm 0.022_{\mathrm{stat}}{}^{+0.050}_{-0.059\,\mathrm{syst}}$.
The hadron-cluster background is caused by a beam-halo neutron that directly hits CSI and produces two hadronic clusters through hadronic interactions. Such clusters are suppressed by exploiting the short radiation length of photons compared with the long interaction length of neutrons: the timing difference between the MPPC and PMT readout of each crystal measures the shower depth, while a convolutional neural network and a Fourier frequency analysis use the cluster and pulse shapes. The number of hadron-cluster background events was estimated to be $0.024\pm0.004_{\mathrm{stat}} \pm 0.006_{\mathrm{syst}}$ using neutron control samples.
An $\eta$ produced by a beam-halo neutron interacting with CV may also become a background, because the reconstructed $Z_{\mathrm{vtx}}$ is shifted upstream into the signal region due to the $\pi^0$ mass assumption in the reconstruction. The $\eta$ production in CV was estimated from MC simulation of beam-halo neutrons. The number of CV-$\eta$ background events was estimated to be $0.023 \pm 0.010_{\mathrm{stat}} \pm 0.005_{\mathrm{syst}}$.
A $\pi^0$ produced via the interaction of a beam-halo neutron with NCC could also become a background, called the upstream $\pi^0$ background. If the photon energy was mismeasured due to photo-nuclear interactions in CSI, the reconstructed $Z_{\mathrm{vtx}}$ could be shifted into the signal region. The upstream $\pi^0$ events were estimated from MC simulation of beam-halo neutrons normalized to data. The number of upstream $\pi^0$ background events was estimated to be $0.060\pm 0.046_{\mathrm{stat}}\pm 0.007_{\mathrm{syst}}$.

\subsection{Sigle event sensitivity}

The single event sensitivity (SES) was obtained from Eq.~\ref{ses}:
\begin{equation}
\mathrm{SES} = \frac{1}{A_{\mathrm{sig}}}\frac{A_{\mathrm{norm}} \cdot \mathrm{BR}(K_L\rightarrow 2\pi^0)}{N_{\mathrm{norm}}},
\label{ses}
\end{equation}
where $A_{\mathrm{norm}}$ ($A_{\mathrm{sig}}$) is the acceptance of $K_L\rightarrow 2\pi^0$ ($K_L\rightarrow \pi^0\nu\bar\nu$) decays evaluated from MC simulations, $\mathrm{BR}(K_L\rightarrow 2\pi^0)$ is the nominal branching fraction \cite{pdg}, and $N_{\mathrm{norm}}$ is the number of reconstructed $K_L\rightarrow 2\pi^0$ events, corrected for the trigger prescale factor. With $A_{\mathrm{sig}} = 0.475\%$, $A_{\mathrm{norm}} = 0.334\%$ and $N_{\mathrm{norm}} = 6.52 \times 10^5$, the SES on the 2021 data was estimated to be $(9.33\pm 0.06_{\mathrm{stat}} \pm 0.84_{\mathrm{syst}})\times 10^{-10}$.
The systematic uncertainties on the SES are summarized in the right table in Fig.~\ref{tables}. The major contributions were from the kinematic cuts for $K_L\rightarrow 2\pi^0$ (4.6\%), shape-related cuts (4.2\%), kinematic cuts for $K_L\rightarrow \pi^0\nu\bar\nu$ (3.5\%), veto cuts (3.4\%), and normalization modes inconsistency (3.5\%), for a total relative uncertainty of 9.0\%. The uncertainties other than the last one were estimated by evaluating the acceptance discrepancy between data and MC for each cut and summing the differences quadratically.

\begin{figure}[htb]
    \begin{center}
        {\includegraphics[width=.44\linewidth]{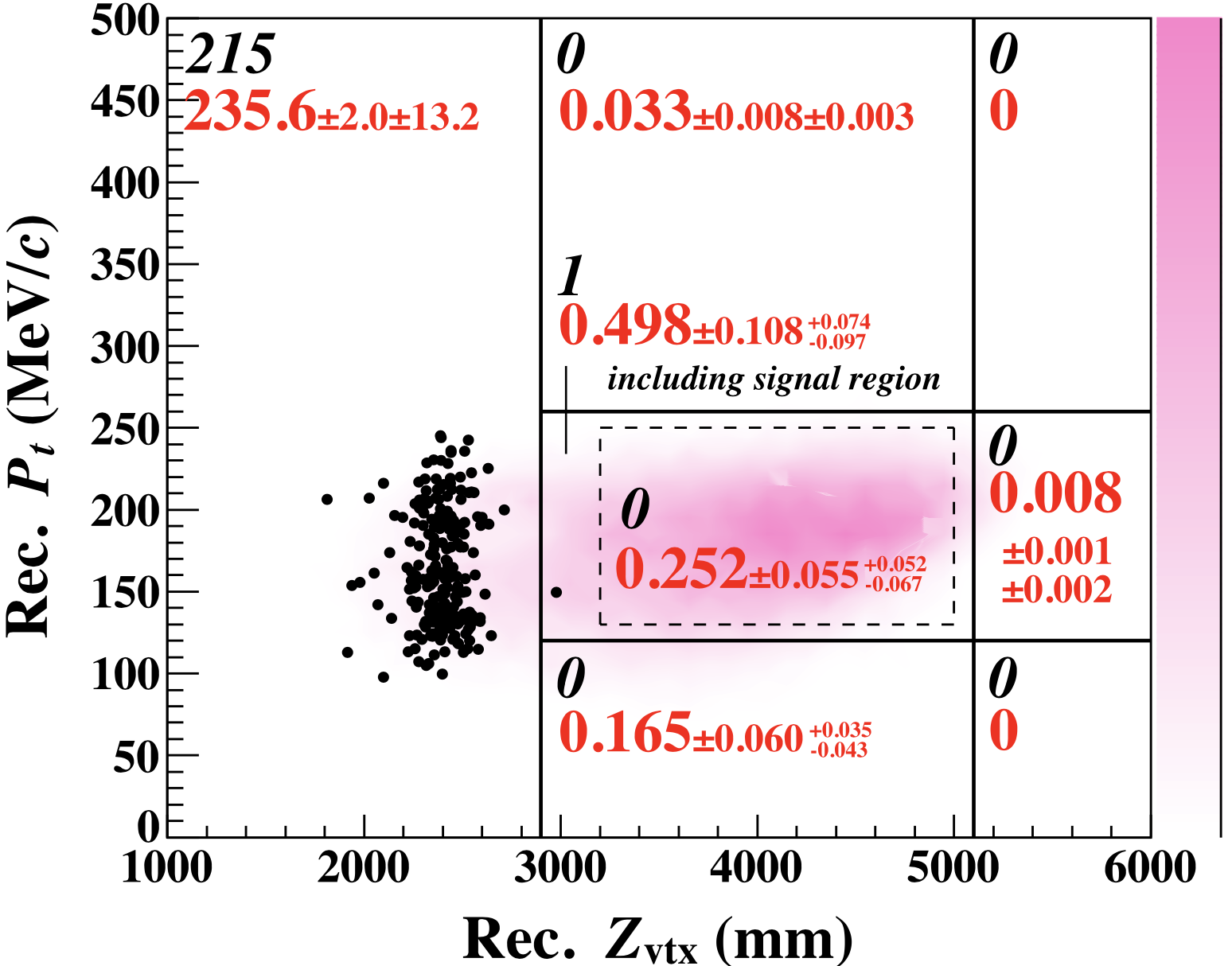}}\hspace{0.5cm}
        {\includegraphics[width=.49\linewidth]{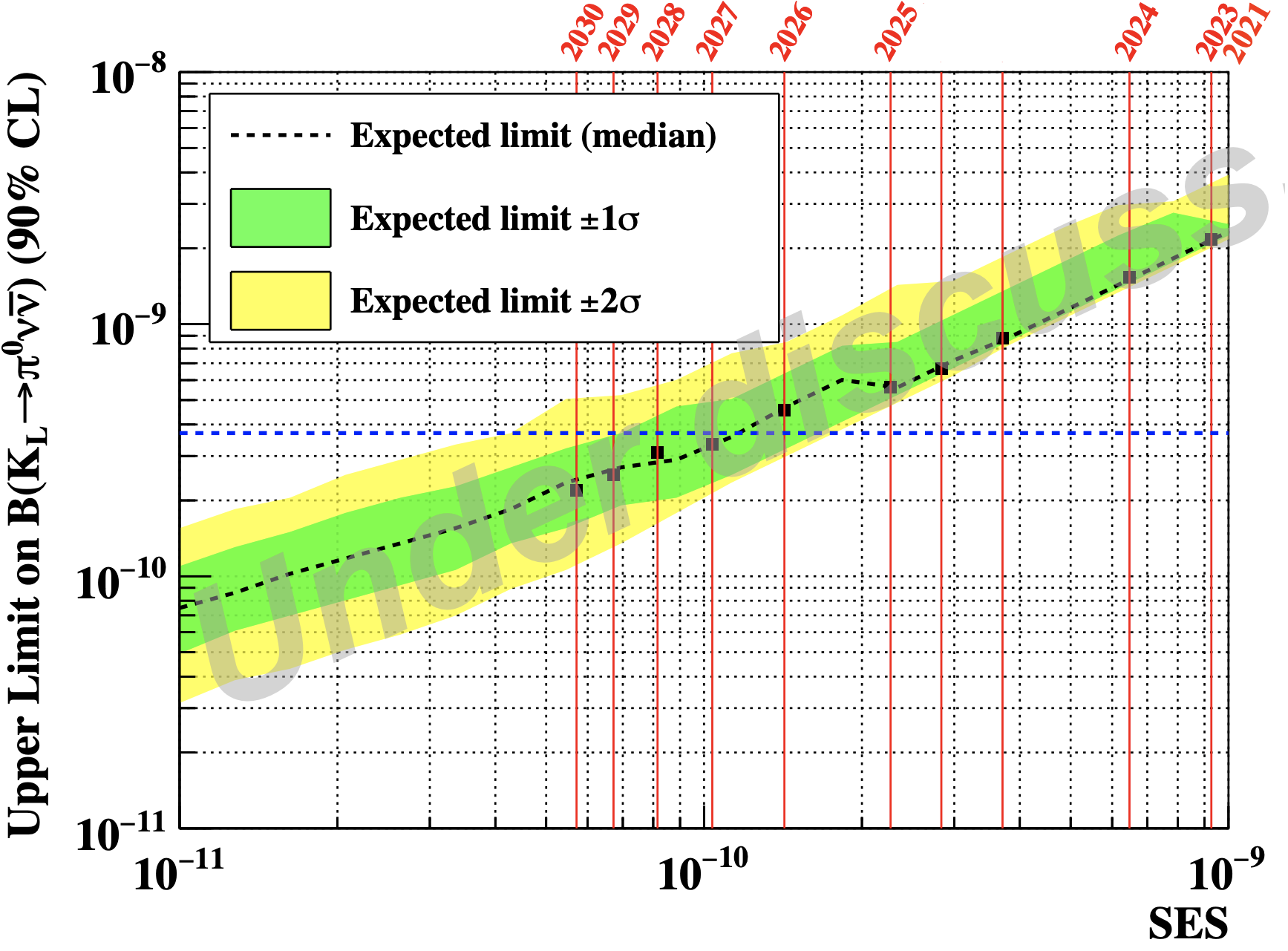}}
        \caption{\it Left: reconstructed $\pi^0$ transverse momentum ($P_t$) versus decay vertex position ($Z_{\mathrm{vtx}}$) of the events surviving the signal selection. The signal region is surrounded by dashed lines. Black dots: observed events; shaded contour: $K_L \rightarrow \pi^0\nu\bar\nu$ MC distribution. Black italic (red regular) numbers: observed (expected background) events in each region; the second and third red numbers are the statistical and systematic uncertainties. Right: projected expected limit versus SES under the stated assumptions, including the statistics already collected and 3 months per year of data taking up to 2030.}
        \label{result}
    \end{center}
\end{figure}

\subsection{Analysis result and prospects}
The events surviving all selection criteria in the 2021 data are shown in Fig.~\ref{result} (left), together with the expected background in each control region. No events were found in the signal region after unblinding. Using Poisson statistics and taking the uncertainty on the SES into account \cite{set_ul}, the resulting upper limit is $\mathrm{BR}(K_L\rightarrow \pi^0\nu\bar\nu) < 2.2 \times 10^{-9}$ at 90\% C.L.

KOTO expects to take data until 2030. Assuming that the beam power will reach 100 kW after 2026, with 20 days/month of beam time, 3 months/year of data taking and 70\% efficiency for physics data taking, the experiment aims to reach $\mathrm{SES} < 10^{-10}$ in 3-4 years. The projected upper limit is shown in Fig.~\ref{result} (right).

\section{KOTO II experiment}

A next-generation experiment, KOTO II \cite{proposal}, is planned to collect tens of signal events, aiming at the measurement of the $K_L \rightarrow \pi^0\nu\bar\nu$ branching ratio through the observation of the decay.

\subsection{Beam line}
To achieve high experimental sensitivity for the $K_L \rightarrow \pi^0\nu\bar\nu$ measurement, the $K_L$ flux, the signal acceptance, and the signal-to-background ratio must be maximized. The $K_L$ flux is determined by the production angle and the solid angle of the secondary neutral beam; other important parameters are the primary proton beam intensity and the target properties. The $K_L$ production angle is 5 degrees at KOTO II instead of 16 degrees at KOTO, chosen to obtain a larger $K_L$ flux and a harder $K_L$ momentum spectrum while limiting the neutron fraction. The solid angle, defined by a collimator, is $4.8~\mu$sr. This beam requires a new experimental area behind the primary beam dump and a new target station close to the dump.

\begin{figure}[htb]
    \begin{center}
        {\includegraphics[width=1\textwidth]{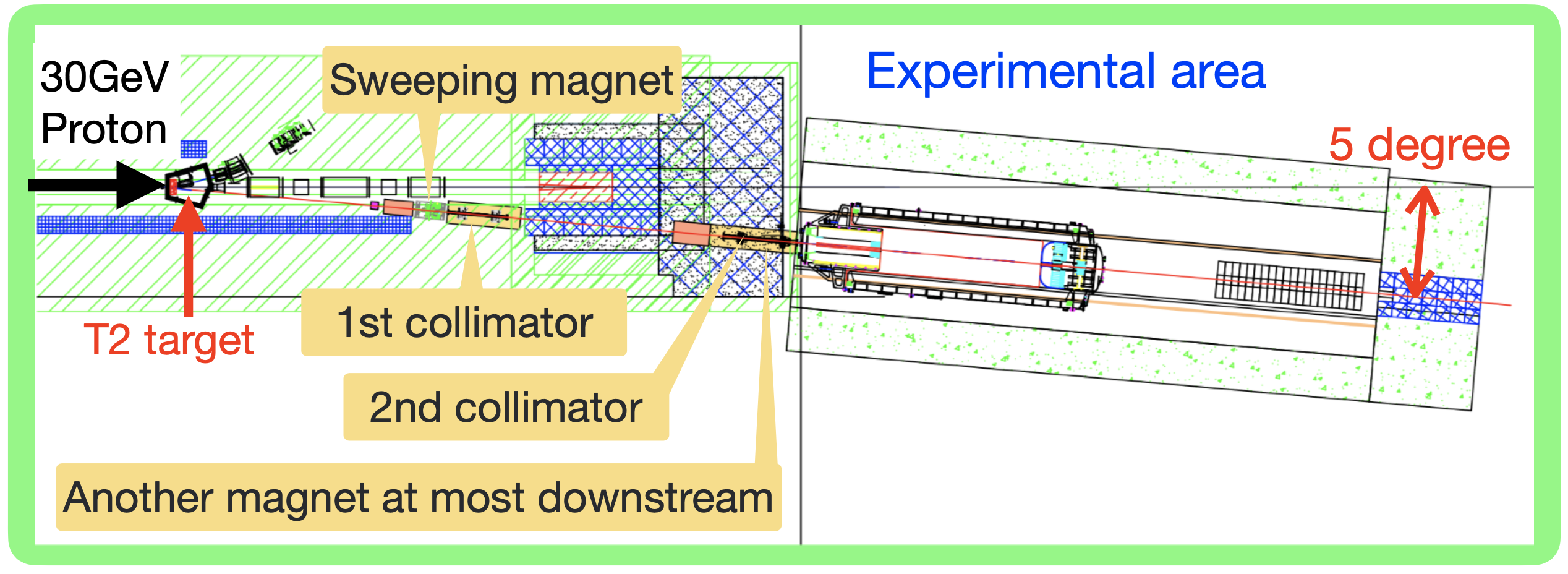}}
        \caption{\it Schematic drawing of the KOTO II beam line in the Extended Hadron Experimental Facility. The experimental area is located behind the beam dump. Distance between the T2 target and the experimental area is assumed to be 43 m in the present studies. A modeled cylindrical detector with 20 m in length and 3 m in diameter is shown in the experimental area as a reference.}
\label{beam_line}
    \end{center}
\end{figure}

Fig.~\ref{beam_line} shows a possible configuration with a new target (T2). The experimental area behind the dump is 43 m from T2, compared to 20 m in KOTO. The Hadron Experimental Facility (Hadron Hall) must therefore be expanded to approximately double its experimental area. The $K_L$ momentum spectrum at 43 m from T2 peaks at 3 GeV$/c$. The $K_L$ yield is $5.5 \times 10^7$ per $10^{14}$ protons on target (POT). For a 100 kW beam with 2 s spills every 4.2 s ($\sim 10^{14}$ POT per spill), the incident rates at the detector are 24 MHz for $K_L$'s, 450 MHz for neutrons with energy above 1 GeV, and 110 MHz for photons with energy above 10 MeV.

\subsection{KOTO II detector}

\begin{figure}[htb]
    \begin{center}
        {\includegraphics[width=0.9\textwidth]{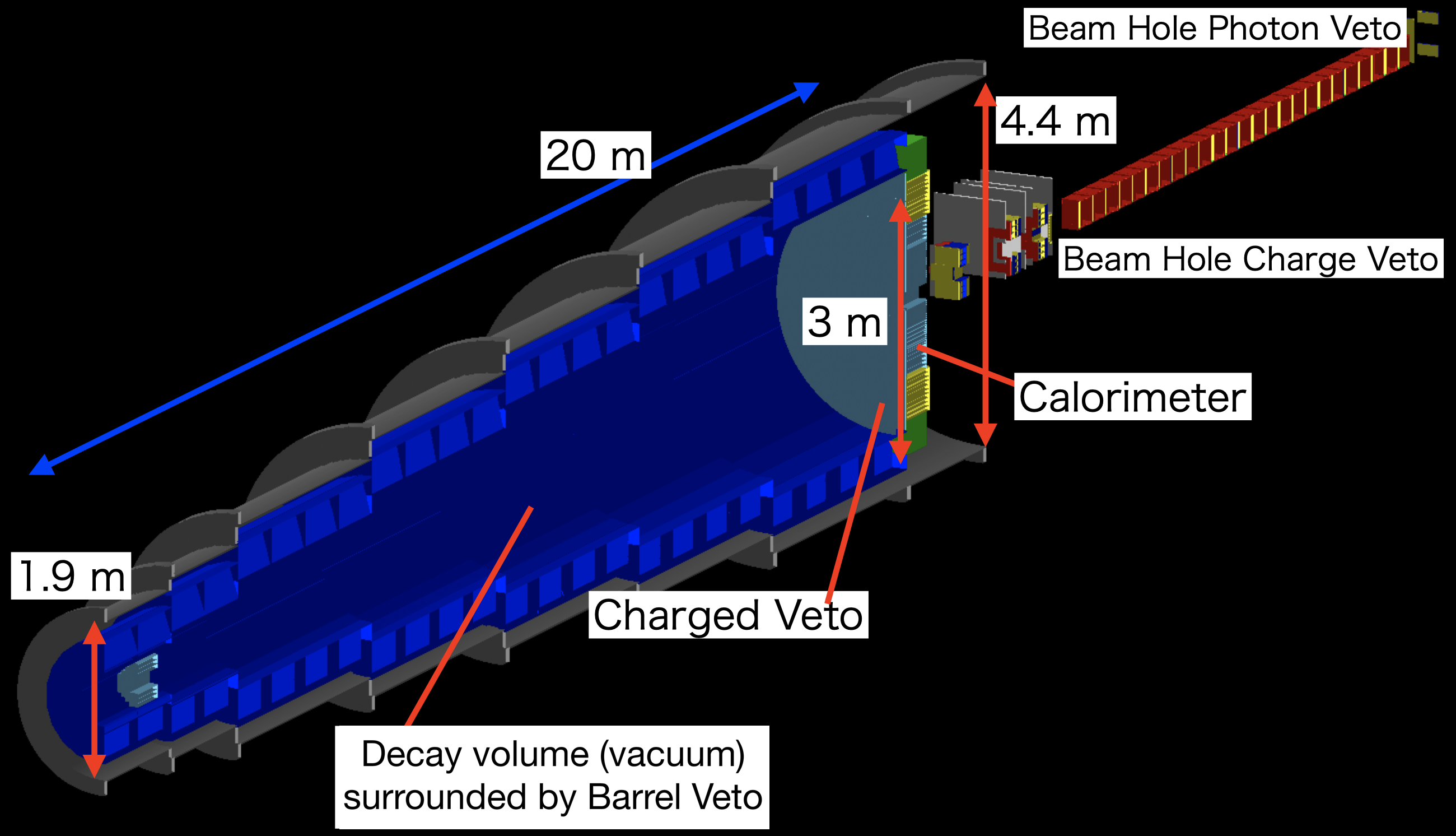}}
        \caption{\it Schematic view of the KOTO II detector. The beam travels along the axis from left to right; the $z$ axis points downstream, with the origin at the entrance of the detector.}
        \label{koto2_schematic}
    \end{center}
\end{figure}

The conceptual base design of the KOTO II detector is shown in Fig.~\ref{koto2_schematic}. The signal region is $3 < z < 15$ m. The detector is cylindrical, with its axis along the beam; the calorimeter is 3 m in diameter at $z = 20$ m, and veto counters surround the signal region.

The KOTO II Collaboration is developing numerous innovative detector technologies. The main calorimeter may use the undoped CsI crystals of KOTO for the inner part, surrounded by a shashlyk calorimeter (a sandwich of lead and plastic scintillator plates with wavelength-shifting fibers). For the Barrel Veto, ring-type counters made of lead and plastic scintillator plates stacked in the $z$ direction are under discussion. Thin B$_4$C films between layers are foreseen to absorb thermal neutrons and reduce accidental hits, exploiting the large thermal-neutron capture cross section of $^{10}$B. To further reduce the background, a granular preshower to measure the photon incidence angle is under study: a finely segmented calorimeter of five radiation lengths with 15 mm strips read out in $x/y$, placed in front of the main calorimeter, can provide an angular resolution of 1-3 degrees for photons of several hundred MeV, which is expected to substantially reduce the backgrounds induced by wrongly reconstructed decay vertices, such as beam-halo $K_L\rightarrow 2\gamma$ decays.
The Beam Hole Charged and Photon Vetoes (BHCV and BHPV) operate inside the neutral beam, where the extremely high rates of beam photons and neutrons make them particularly demanding. They must be as insensitive as possible to $n$ and low-energy $\gamma$ to minimize accidental coincidences (random vetoes) without losing efficiency. For the BHCV, a silicon pixel detector with 2 layers of 200 $\mu$m thickness and a 5 ns time window is under consideration, which would reduce the random veto from the present 8\% in KOTO to below 4\%. For the BHPV, an ultra-fast, highly segmented PWO-UF crystal calorimeter, developed for the HIKE project, is under consideration. Its fast timing allows a shorter veto window, thus reducing signal loss from accidental hits. The foreseen time resolution is $\sigma_t \sim 100$ ps, compared with $\sim 600$ ps for the BHPV of KOTO.

\subsection{Signal selection and expected sensitivity}

\begin{figure}[htb]
    \begin{center}
        {\includegraphics[width=.48\linewidth]{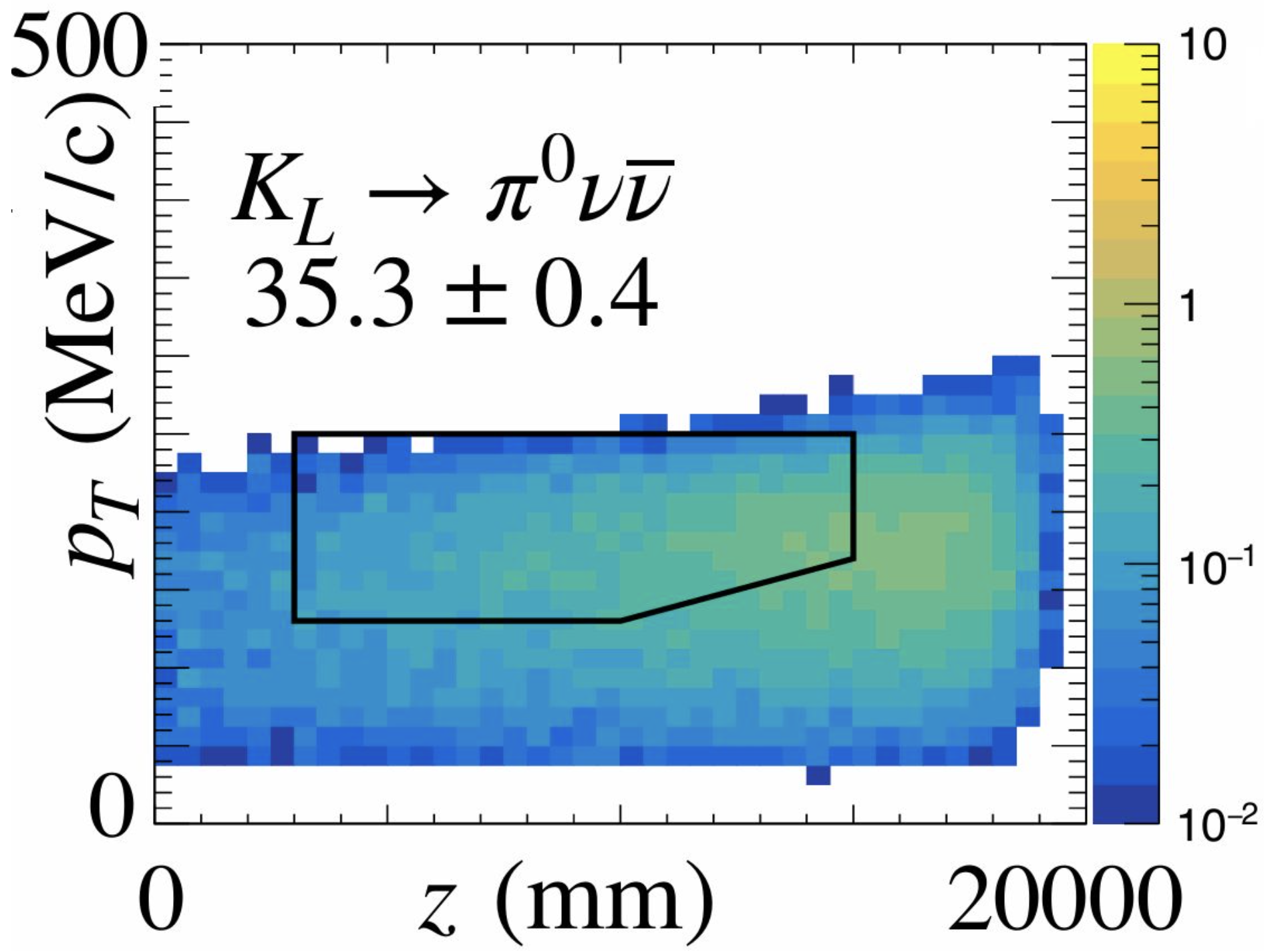}}\hspace{0.5cm}
        {\includegraphics[width=.44\linewidth]{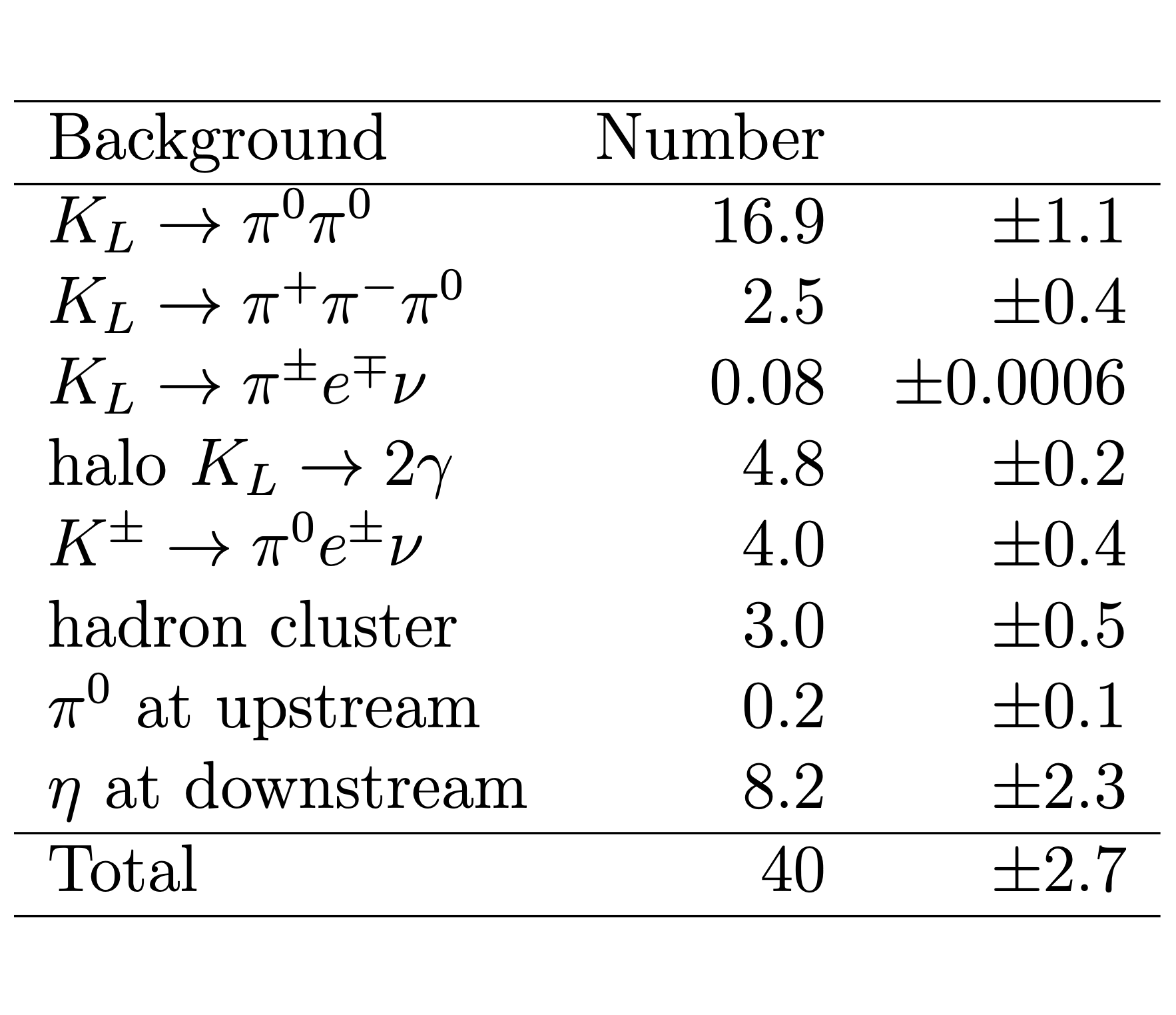}}
        \caption{\it Left: $P_t$ vs $Z_{\mathrm{vtx}}$ distribution of the $K_L \rightarrow \pi^0\nu\bar\nu$ signal and expected number of events for a running time of $3 \times 10^7$ s. Right: summary of the KOTO II background estimation.}
        \label{simulation}
    \end{center}
\end{figure}

The signal selection requires two photons from the $\pi^0$ decay reconstructed in the calorimeter and no other hits in the calorimeter or the veto counters, since there are no other visible particles in the decay. The vertex position ($Z_{\mathrm{vtx}}$) of the $\pi^0$ decay is reconstructed assuming that it lies on the beam axis and that the two-photon invariant mass equals the nominal $\pi^0$ mass. The transverse momentum ($P_t$) of the $\pi^0$ is then obtained from $Z_{\mathrm{vtx}}$; a large $P_t$ within a given $Z_{\mathrm{vtx}}$ region is required, since the two neutrinos carry away momentum. The simulated distribution of events in the $P_t$ vs $Z_{\mathrm{vtx}}$ plane is shown in Fig.~\ref{simulation} (left).

\begin{figure}[htb]
    \begin{center}
        {\includegraphics[width=0.8\textwidth]{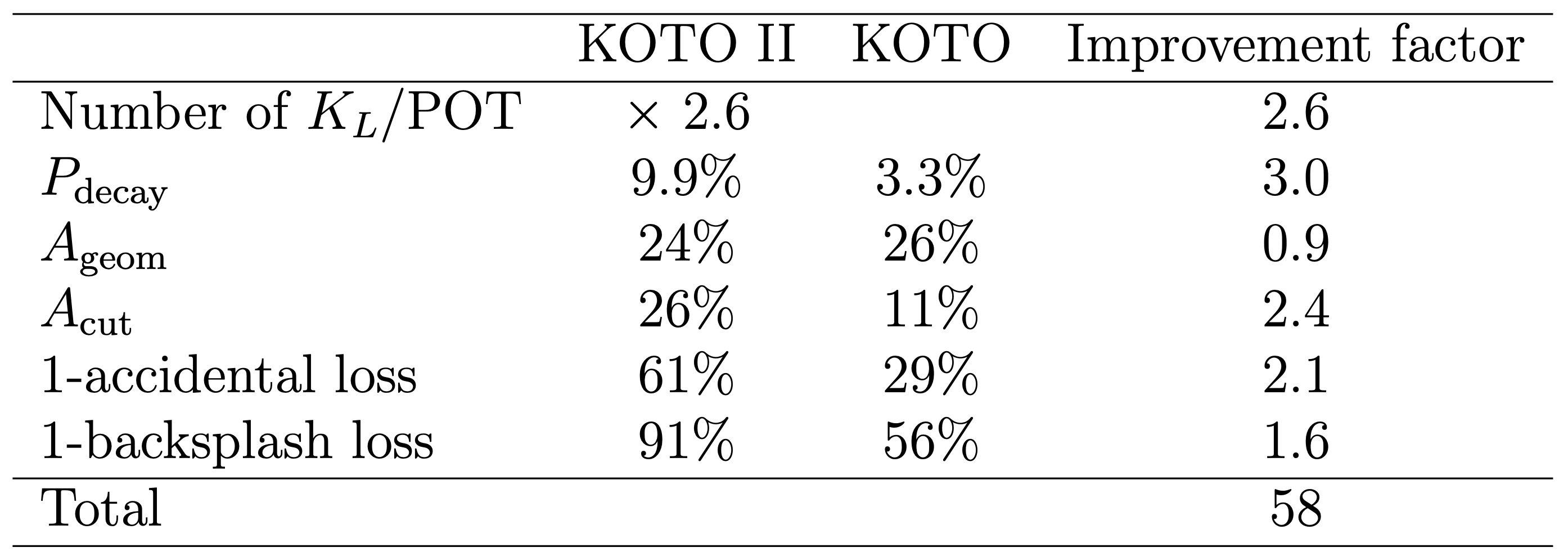}}
        \caption{\it Comparison of $K_L$ yield and signal acceptances between KOTO and KOTO II.}
        \label{tabella_valori}
    \end{center}
\end{figure}

A running time of $3 \times 10^7$ s (corresponding to about 5 years of 3 months per year of data taking at 70\% efficiency) with a 100 kW beam is considered.
The signal yield is evaluated from the $K_L$ flux, the $K_L$ decay probability in the signal $z$ region, the acceptance of two photon hits in the calorimeter, and the event selection acceptance. Signal losses from accidental hits and shower backsplash in the veto counters must also be considered. The values of these variables from the KOTO II simulation are shown in the table in Fig.~\ref{tabella_valori}, together with those of KOTO.

The expected number of signal events in $3 \times 10^7$ s of running, assuming the SM $\mathrm{BR}(K_L \rightarrow \pi^0\nu\bar\nu) \simeq 3 \times 10^{-11}$, is $35$. For the same $K_L$ flux, the expected backgrounds from simulation are listed in the table in Fig.~\ref{simulation} (right). A total of 40 background events is expected, dominated by $K_L\rightarrow \pi^0\pi^0$, giving a signal-to-background ratio (S/B) of 0.89. Assuming the SM branching ratio, an observation at the $5.6 \sigma$ level is expected, with a statistical precision of about 25\% on $\mathrm{BR}(K_L \rightarrow \pi^0\nu\bar\nu)$.

\subsection{Measurement of $\mathrm{BR}(K_L\rightarrow \pi^0 \ell^+ \ell^-)$}

The ultra-rare $K_L\rightarrow \pi^0 \ell^+ \ell^-$ decays ($\ell = e,\mu$) are theoretically clean golden modes in kaon physics, allowing the exploration of new physics (NP) contributions in the $s \rightarrow d\ell\ell$ short-distance interaction, to be compared with the $b \rightarrow s\ell\ell$ transition \cite{leptons}. A measurement of the $K_L \rightarrow \pi^0 e^+ e^-$ decay rate with 25\% precision would provide crucial complementary information for the interpretation of the $K \rightarrow \pi\nu\bar\nu$ measurements in terms of NP models with lepton flavour universality (LFU) violation.
Its measurement is the primary target of a second phase of KOTO II, in which the detector setup will be optimized with the addition of charged-particle tracking.

\section{Conclusion}

KOTO is expected to reach a sensitivity below $10^{-10}$ in 3--4 years, but observing $K_L\rightarrow \pi^0\nu\bar\nu$ at the SM rate requires a next-generation experiment. KOTO II is designed to collect about 35 signal events over 40 background events in 5 years, allowing a $5.6\sigma$ observation and a branching ratio measurement with 25\% statistical precision. Its second phase, with charged-particle tracking, will extend the program to $K_L\rightarrow \pi^0\ell^+\ell^-$, other rare decays and hidden-sector searches. 
KOTO II has received J-PARC Stage-1 approval; the collaboration has been formed, R\&D is finalizing the detector design, and a TDR towards Stage-2 approval is in preparation. The extension of the Hadron Experimental Facility, a top-priority project of KEK, will enable the realization of KOTO II, and its funding is being actively pursued.

\end{document}